\documentclass{natureprintstyle}
\usepackage{graphicx}
\usepackage{color}
\usepackage{hyperref}
\usepackage{longtable}
\usepackage{amssymb}

\usepackage{lineno}
\usepackage{caption}
\def\deg{^\circ}
 
\def\kpc{{\rm\,kpc}}
\def\kms{{\rm\,km\,s^{-1}}}
\def\msun{{\rm\,M_\odot}}

\def\pc{{\rm\,pc}}

\def\lsun{{\rm L}_\odot}

\def\ferre{{\tt FERRE\,\,}}

\def\lta{\mathrel{\spose{\lower 3pt\hbox{$\mathchar"218$}}
     \raise 2.0pt\hbox{$\mathchar"13C$}}}
\def\gta{\mathrel{\spose{\lower 3pt\hbox{$\mathchar"218$}}
     \raise 2.0pt\hbox{$\mathchar"13E$}}}

     \def\Gyr{{\rm\,Gyr}}

\def\FeH{{\rm[Fe/H]}}

\usepackage{tikz}

\newcommand{\aap}{Astron. Astrophys.}
\newcommand{\araa}{A. R. of Astron. Astrophys.}

\newcommand{\apj}{Astrophys. J.}
\newcommand{\apjs}{Astrophys. J. Supp.}
\newcommand{\apjl}{Astrophys. J. Letters}
\newcommand{\aj}{Astronom. J.}
\newcommand{\mnras}{Mon. Not. R. Astron. Soc.}
\newcommand{\nat}{Nature}
\newcommand{\aapr}{Astron. Astrophys. R.}

\newcommand{\aaps}{A\&ASupp}

\title{A stellar stream remnant of a globular cluster below the metallicity floor}

\author{Nicolas F. Martin$^{*1,2}$, Kim A. Venn$^{3}$, David S. Aguado$^{4,5,6}$, Else Starkenburg$^7$, Jonay I. Gonz\'alez Hern\'andez$^{5,8}$, Rodrigo A. Ibata$^1$, Piercarlo Bonifacio$^9$, Elisabetta Caffau$^9$, Federico Sestito$^3$, Anke Arentsen$^1$, Carlos Allende Prieto$^{5,8}$, Raymond G. Carlberg$^{10}$, S\'ebastien Fabbro$^{3,11}$, Morgan Fouesneau$^2$, Vanessa Hill$^{12}$, Pascale Jablonka$^{13,9}$, Georges Kordopatis$^{12}$, Carmela Lardo$^{14}$, Khyati Malhan$^{15}$, Lyudmila I. Mashonkina$^{16}$, Alan W. McConnachie$^{11}$, Julio F. Navarro$^{3}$, Rub\'en S\'anchez Janssen$^{17}$, Guillaume F. Thomas$^{5,8}$, Zhen Yuan$^1$, Alessio Mucciarelli$^{14,18}$}

\begin{document}

\maketitle

\begin{affiliations}
\item Universit\'e de Strasbourg, CNRS, Observatoire astronomique de Strasbourg, UMR 7550, F-67000, France
\item Max-Planck-Institut f\"ur Astronomie, K\"onigstuhl 17, D-69117, Heidelberg, Germany
\item Department of Physics and Astronomy, University of Victoria, PO Box 3055, STN CSC, Victoria BC V8W 3P6, Canada
\item Institute of Astronomy, University of Cambridge, Madingley Road, Cambridge CB3 0HA, UK
\item Instituto de Astrof\'isica de Canarias, V\'ia L\'actea, 38205 La Laguna, Tenerife, Spain
\item Dipartimento di Fisica e Astronomia, Universit\'a degli Studi di Firenze, Via G. Sansone 1, I-50019 Sesto Fiorentino, Italy
\item Kapteyn Astronomical Institute, University of Groningen, Landleven 12, NL-9747AD Groningen, the Netherlands
\item Universidad de La Laguna, Departamento de Astrof\'isica, 38206 La Laguna, Tenerife, Spain
\item GEPI, Observatoire de Paris, Universit\'e PSL, CNRS, 5 Place Jules Janssen, 92195, Meudon, France
\item Department of Astronomy \& Astrophysics, University of Toronto, Toronto, ON M5S 3H4, Canada
\item NRC Herzberg Astronomy \& Astrophysics, 5071 West Saanich Road, Victoria, British Columbia, Canada V9E 2E7
\item Universit\'e C\^ote d'Azur, Observatoire de la C\^ote d'Azur, CNRS, Laboratoire Lagrange, Nice, France
\item Laboratoire d'astrophysique, \'Ecole Polytechnique F\'ed\'erale de Lausanne (EPFL), Observatoire, 1290 Versoix, Switzerland
\item Dipartimento di Fisica e Astronomia, Universit\`a degli Studi di Bologna, Via Gobetti 93/2, I-40129 Bologna, Italy
\item The Oskar Klein Centre, Department of Physics, Stockholm University, AlbaNova, SE-10691 Stockholm, Sweden
\item Institute of Astronomy, Russian Academy of Sciences, RU-119017 Moscow, Russia
\item UK Astronomy Technology Centre, Royal Observatory, Blackford Hill, Edinburgh, EH9 3HJ, UK
\item INAF - Osservatorio di Astrofisica e Scienza dello Spazio di Bologna, Via Gobetti 93/3, I-40129 Bologna, Italy

\end{affiliations}

\begin{abstract}

Stellar ejecta gradually enrich the gas out of which subsequent stars form, making the least chemically enriched stellar systems direct fossils of structures formed in the early universe\cite{frebel15}. Although a few hundred stars with metal content below one thousandth of the solar iron content are known in the Galaxy\cite{yong13,li18b,aguado19}, none of them inhabit globular clusters, some of the oldest known stellar structures. These show metal content of at least $\sim0.2$ percent of the solar metallicity ($\FeH \gtrsim -2.7$). This metallicity floor appears universal\cite{beasley19,wan20} and it has been proposed that proto-galaxies that merge into the galaxies we observe today were simply not massive enough to form clusters that survived to the present day\cite{kruijssen19}. Here, we report the discovery of a stellar stream, C-19, whose metallicity is less than 0.05 per cent the solar metallicity ($\FeH=-3.38\pm0.06\textrm{ (stat.)} \pm0.20\textrm{ (syst.)}$). The low metallicity dispersion and the chemical abundances of the C-19 stars show that this stream is the tidal remnant of the most metal-poor globular cluster ever discovered, and significantly below the purported metallicity floor: clusters with significantly lower metallicities than observed today existed in the past and contributed their stars to the Milky Way halo.
\end{abstract}

\captionsetup[figure]{labelfont={bf}}
\begin{figure}
\begin{center}
\includegraphics[width=1.05\hsize,angle=0]{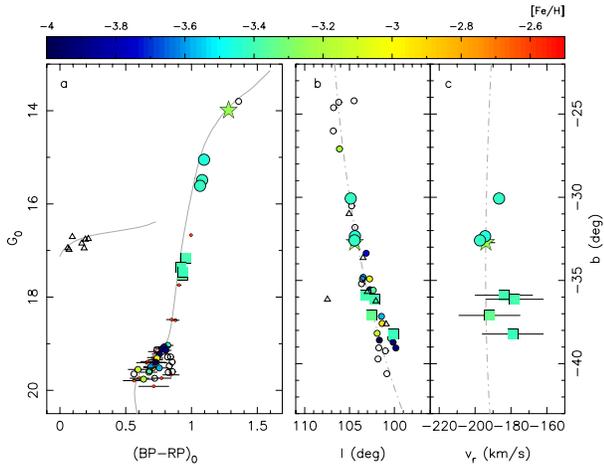}
\caption{\footnotesize Properties of the C-19 member stars. \emph{(a)} Colour-magnitude diagram of the C-19 stream. Stars are colour-coded by their photometric metallicity as measured in the \emph{Pristine} survey (small symbols), or their spectroscopic metallicity (large symbols) from spectra obtained with LAMOST (star symbol), Gemini/GRACES (circles), or GTC/OSIRIS (squares). The \emph{Pristine} metallicities unambiguously show that C-19 is very metal-poor and helps to remove the few contaminants with higher metallicities ($\FeH_\mathrm{Pristine}>-2.5$ shown with red, smaller symbols and removed in the other panels; see Methods). Hollow circles denote stars with no metallicity measurement but that have a proper motion consistent with the C-19 stream in this region of the sky. These stars show the clear and very blue horizontal branch of the system and also include a few likely members along the red-giant branch. The grey track is a 13-Gyr PARSEC isochrone with the lowest available metallicity ([M/H] $=-2.2$), at the favored heliocentric distance of 18\,kpc. Despite a metallicity that is significantly more metal-rich than the spectroscopic metallicity of C-19, this isochrone provides a good match to the member stars because of the low sensitivity of the Gaia colour to metallicity changes in this metal-poor regime. \emph{(b)} Distribution of C-19 candidate member stars on the sky, in Galactic coordinates. A blue horizontal branch star candidate, located at $(\ell,b)=(94,6\deg,-52.7\deg)$ and significantly detached from the rest of the stream stars, is not shown here but could indicate that C-19 is at least $15\deg$ longer than shown here. \emph{(c)} Radial velocities, $v_\mathrm{r}$, of confirmed C-19 stars with spectroscopy. In the two right-most panels, the dot-dashed grey lines represent the calculated track of the C-19 orbit (see Methods). In all panels, the error bars represent the uncertainties, calculated as $\pm1\sigma$ Gaussian standard deviations.\label{CMD}}
\end{center}
\end{figure}

The C-19 stellar stream is a grouping of stars in the Milky Way halo that share a common orbital motion around the Galaxy and was discovered through the application of the \texttt{STREAMFINDER} algorithm\cite{ibata21} to the astrometric data of the Gaia Early Data Release 3\cite{Gaia21a}. The stellar structure has a very low density and extends over $\sim15\deg$ on the sky. The reality of the stellar stream was confirmed via the photometric metallicities (solar-scaled iron content, \FeH) of the \emph{Pristine} survey\cite{starkenburg17b}, which revealed the coherent and extremely low metallicity of candidate member stars. Especially for the brighter stars, a regime with little expected contamination, the Pristine photometric metallicities exhibit a very coherent metallicity at an extremely metal-poor level. Selecting stars with Gaia proper motions similar to those of the \texttt{STREAMFINDER} sample yields the colour-magnitude diagram (CMD) of stars associated with the structure at high significance, shown in Figure~\ref{CMD}. The CMD of the stream displays a sparsely populated but well-defined red-giant branch that extends into a main-sequence turnoff at the faint end of the data, along with a clear, and very blue, horizontal branch. These features are typical of old and metal-poor globular clusters. Correcting for stars that are fainter than the magnitude limit of the sample, we estimate that the total luminosity and stellar mass of the C-19 progenitor is at least $3.5\times10^3\lsun$ and $0.8\times10^4\msun$, respectively, at the favored heliocentric distance of 18 kpc (see Methods). This can only be a lower limit as the C-19 stream may extend over a larger region of the sky but with lower densities that prevent its detection at this stage. This mass is however quite typical for halo globular clusters\cite{harris96}.

We obtained spectroscopic observations of eight member stars with two telescopes and spectrographs (Gemini/GRACES and GTC/OSIRIS) to determine the nature of the C-19 progenitor and refine the orbit of the stream (see Methods). The brightest of those stars was also observed and analyzed in the very metal-poor star sample of LAMOST\cite{li18b}. The spectra of all eight of our \emph{Pristine}-selected very metal-poor member stars ($\FeH<-2.0$) yield radial velocities that are in agreement with the predictions from \texttt{STREAMFINDER} and therefore confirm that they are stream members. A ninth star with a significantly more metal-rich \emph{Pristine} metallicity has a discrepant radial velocity and is confirmed to be an outlier. The kinematics of the eight member stars show that C-19 follows an orbit deep in the potential well of the Milky Way, with a pericenter of $\sim7\kpc$ and an apocenter of $\sim27\kpc$, on a plane that is almost polar with respect to the Milky Way plane (see Methods).

\captionsetup[figure]{labelfont={bf}}
\begin{figure*}
\begin{center}
\includegraphics[width=\hsize]{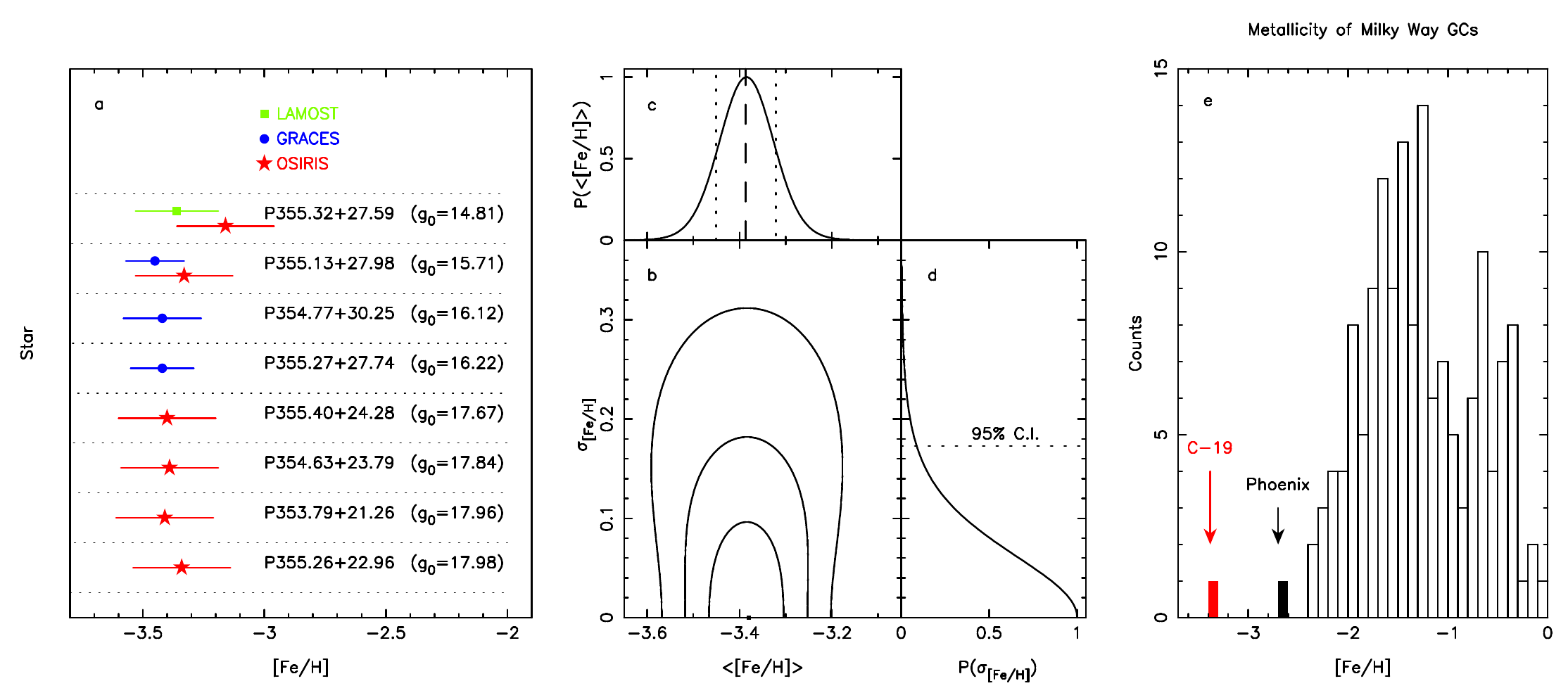}
\caption{\small Metallicity properties of C-19 and its stars observed with spectroscopy. \emph{(a)} Spectroscopic metallicities [Fe/H] calculated for 8 members of C-19. The different colours and symbols denote spectra obtained with the different telescopes, instruments, and/or surveys listed at the top of the panel. The two brightest stars were observed with two different facilities each. \emph{(b)} Probability distribution functions of the mean and dispersion of the metallicities of C-19 stars, assuming they follow a Gaussian distribution. The individual uncertainties in the measurements are taken into account and stars observed twice had their measurements combined, weighted by their individual uncertainties. The marginalized probability distribution functions for the two parameters are shown in panels (c) and (d). The dotted lines in panel (c) represents the uncertainties (Gaussian standard deviation) on the derived metallicity and, in panel (d), it highlights the 95-percent confidence limit. \emph{(e)} Metallicity distribution of all known globular clusters of the Milky Way\cite{harris96}. The metallicity of the C-19 progenitor is highlighted in red, along with the metallicity of the recently discovered Phoenix stream\cite{wan20}, the lowest metallicity stellar stream from a globular cluster previously known to date.\label{FeH}}
\end{center}
\end{figure*}

\captionsetup[figure]{labelfont={bf}}
\begin{figure}
\begin{center}
\includegraphics[width=\hsize]{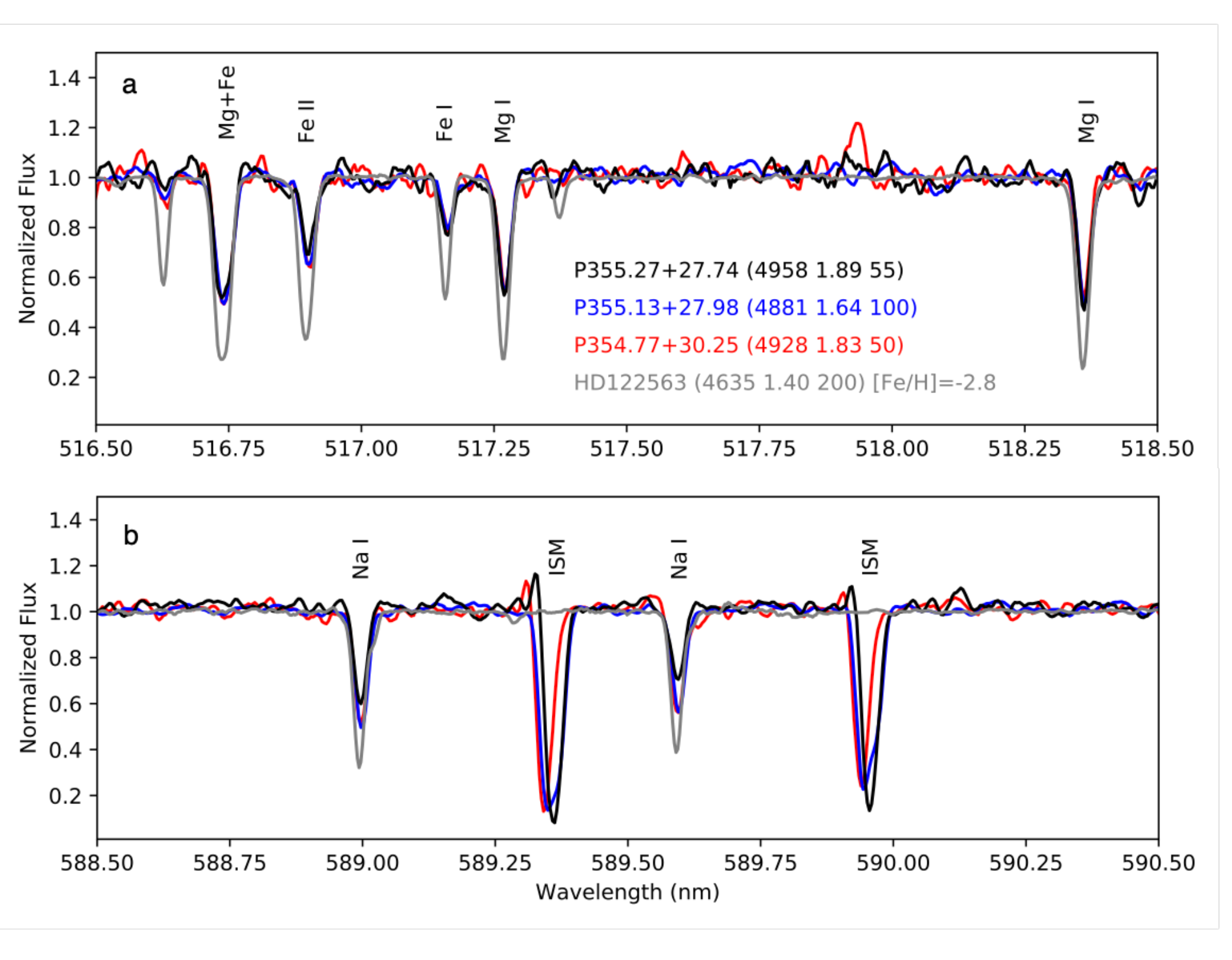}
\caption{\small High resolution GRACES spectra for three members of C-19. All three stars have very similar stellar parameters, showing the similarities in the Mg I and Fe II absorption features, but a significantly lower Na\,I line strength in Pristine\_355.27+27.74. Spectra are labelled by the target name, along with the inferred effective temperature, log surface gravity, and signal-to-noise ratio of the spectrum at 600\,nm. The strong lines labeled ISM in the lower panel correspond to sodium absorption by the interstellar medium and are unrelated to the C-19 stars. The deeper lines of standard star HD~122563 that has similar stellar parameters\cite{kielty21} and $\FeH = -2.8$ confirm that the C-19 stars are significantly more metal-poor.\label{spectra}}
\end{center}
\end{figure}

Spectroscopic metallicities are determined for all eight confirmed C-19 member stars (see Methods). These metallicities are displayed in Fig.~\ref{FeH} and high resolution spectra for three stars from the Gemini/GRACES facility are shown in Fig.~\ref{spectra}. These results unambiguously confirm the extremely low metallicity of C-19, with a mean metallicity $\FeH=-3.38\pm0.06\textrm{ (stat.)}\pm0.20\textrm{ (syst.)}$. Such a low global metallicity has so far never been observed for any stellar system in the Milky Way, its surroundings, or beyond.

We are able to place a stringent limit on the unresolved metallicity dispersion of these eight stars, with $\sigma_\FeH<0.18$ at the 95 per cent confidence level. This strongly suggests that the C-19 progenitor was a globular cluster, and not a dwarf galaxy, since the latter would exhibit a large metallicity dispersion stemming from multiple bursts of star formation\cite{willman12,leaman12}. In addition, the mean metallicity of C-19 is significantly below the mean metallicity expected for dwarf galaxies at this luminosity\cite{kirby13}.

A dispersion in the sodium abundances provides even more compelling evidence that C-19 is a disrupted globular cluster. The high-resolution spectra for three member stars of C-19 are used to determine high-precision chemical abundances (see Methods). We find the [Na/Mg] ratios vary by a factor of three between these stars. These variations are typical of stars in ancient globular clusters\cite{gratton12,bastian18}, most likely due to exposure to high-temperature hydrogen burning, and possibly due to multiple or extended star formation at early epochs. These chemical signatures are rarely seen in Galactic halo stars or in low-luminosity dwarf galaxies\cite{ji20}. These sodium variations can even be seen directly from the difference in Na\,I line strengths in Fig.~\ref{spectra}.
 
For extremely metal-poor stars ($\FeH<-3.0$), the heavy neutron-capture element abundances, such as barium, provide another chemical signature of their origins. At these metallicities, stars that form in low-luminosity dwarf galaxies typically have very low [Ba/Fe] ratios\cite{roederer13}, other than a few which have been enriched by a rare rapid-neutron capture process, such as a compact binary merger\cite{cote19} (e.g., Reticulum II\cite{ji16} or Tucana III\cite{hansen17}). Given the signal-to-noise ratio of the GRACES spectra, we can only measure barium in one star (Pristine\_355.13+27.98), where [Ba/Fe] is near solar, consistent with typical values in globular cluster stars\cite{roederer11}. We can also rule out highly enriched Ba abundances in the two other stars observed with GRACES. Carbon provides another clue to the origins of the C-19 stars, since 40\% of field halo stars in the Galaxy are carbon-rich in the considered metallicity range\cite{yoon18}. None of the six C-19 stars with carbon measurements from our OSIRIS spectra are enhanced ([C/Fe]$<0.70$ in all cases although we do note that the uncertainties are relatively large). While this will have to be confirmed through more accurate measurements, this may further suggests that the C-19 progenitor formed from gas enriched differently from that of typical extremely metal-poor halo stars\cite{norris13}. Overall, the chemical signatures of our sample of stars reinforce our conclusion that the progenitor of the C-19 stream was a globular cluster with a metallicity significantly below the current globular cluster metallicity floor.

The extremely low metallicity of the C-19 progenitor helps reduce tensions in our understanding of the metallicity distribution function of the Milky Way stellar halo. Assuming that the contribution of globular clusters to the Milky Way halo is the same at all metallicities, there is a dearth of globular clusters with [Fe/H]$<-2.5$, with none observed but $\sim5$ expected from the metallicity of halo stars\cite{youakim20}. Our discovery of a cluster stream, significantly below that metallicity floor, in addition to other recently discovered globular cluster stellar streams that are close to the metallicity floor\cite{roederer19,wan20}, shows that more metal-poor clusters formed at early times, but did not survive their tidal interactions with the Milky Way. Intriguingly, the recent realization that a globular cluster with a metallicity $\FeH=-2.9$ currently orbits the Andromeda galaxy\cite{larsen20} could also indicate that such low-metallicity structures may survive until the present day in other environs.

The orbit of C-19 has an apocenter of only $\sim27\kpc$, which suggests a very early accretion onto the Milky Way, when the Milky Way's potential well was shallower than it is now. However, if this is the case, it is puzzling that the C-19 stellar stream can still be identified as a coherent structure to this day. The relatively short orbital time ($<0.5$\,Gyr) and the expected presence of baryonic or dark matter substructures that should further dynamically heat the stellar stream\cite{bonaca19} would all work to disperse the stream into the smooth Milky Way stellar halo. This process may well have already happened and could explain the stream's full width ($\sim600\pc$) and radial velocity dispersion ($\sim7\kms$ from the 3 GRACES stars that have accurate velocities and are located close together); both are significantly larger than what is usually measured for globular cluster streams ($<100\pc$ and 1--3$\kms$, respectively\cite{ibata16,erkal17}). Alternatively, or in addition, the fact that the C-19 orbit crosses the disk multiple time over the last $1\Gyr$ (see Methods) could also have significantly heated the stream. It will be fascinating to explore the processes that could have shielded the C-19 progenitor for long enough so that its stream is still visible today. For example, it may have been born in its own host galaxy, which partly shielded C-19 from destructive interactions with the Milky Way. Hunting for host stars on orbits similar to that of the C-19 stream could help to constrain this scenario.

Finally, the C-19 progenitor provides an intriguing window into the formation of globular clusters at very early times. Galaxies are expected to grow hierarchically, and to enrich gradually as more and more stars form and evolve. The very existence of C-19 proves that globular clusters must have been able to form in the lowest metallicity environments as the first galactic structures were assembling. The time-evolution of scaling relations between the mass of a galaxy and its metallicity, built from galaxy-evolution simulations\cite{ma16,kruijssen19}, implies that the C-19 progenitor must have formed within at most the first 2--3\,Gyr after the formation of the universe. After that period of time, no galaxy is expected to still host sufficiently low metallicity gas to have formed a system like C-19. Furthermore, the same scaling relations between metallicity and stellar mass indicate that potential C-19 hosts should have had extremely low masses ($\lesssim4.5\times10^4\msun$), regardless of formation time\cite{ma16}. Given our mass estimate of $\geq0.8\times10^4\msun$, then the star-formation event that led to the C-19 progenitor must have represented a significant fraction of all stars present in the host galaxy at that time. This is compelling evidence that structures like C-19 provide strong constraints on galaxy and globular cluster formation models at the earliest times.


\newpage

\setcounter{figure}{0}  
\captionsetup[figure]{labelfont={bf},name={Extended Data Figure}}

\begin{center}
{\bf \Large \uppercase{Methods} }
\end{center}

\section{Discovery of the C-19 stream}
The C-19 stream was discovered\cite{ibata21} as a grouping of stars along the same orbit by applying the \texttt{STREAMFINDER} algorithm\cite{malhan18} to the Early Data Release 3 (EDR3) of the European Space Agency's Gaia space mission\cite{Gaia21a}. The list of all those stars is given in Extended Data Table~\ref{tab:members}. The C-19 stars span $\sim15\deg$ on the sky, centered near $(l,b)=(107\deg,-37\deg)$, as shown in the middle panel of Figure~\ref{CMD}. The \texttt{STREAMFINDER} analysis favors a heliocentric distance of $\sim18\kpc$ for this system, confirmed by the combined orbital analysis of the full sample of likely members (see below). As can be seen in Figure~\ref{CMD}, the full width of the stream is $\sim2\deg$, or $\sim600\pc$ at the favored distance. This is large for a typical globular cluster stream, likely indicating that C-19 was heated during its orbiting of the Milky Way and/or that it was already tidally disrupted in its (putative) host dwarf galaxy\cite{carlberg18,malhan19b}.

\begin{figure}
\begin{center}
\includegraphics[width=1.1\hsize]{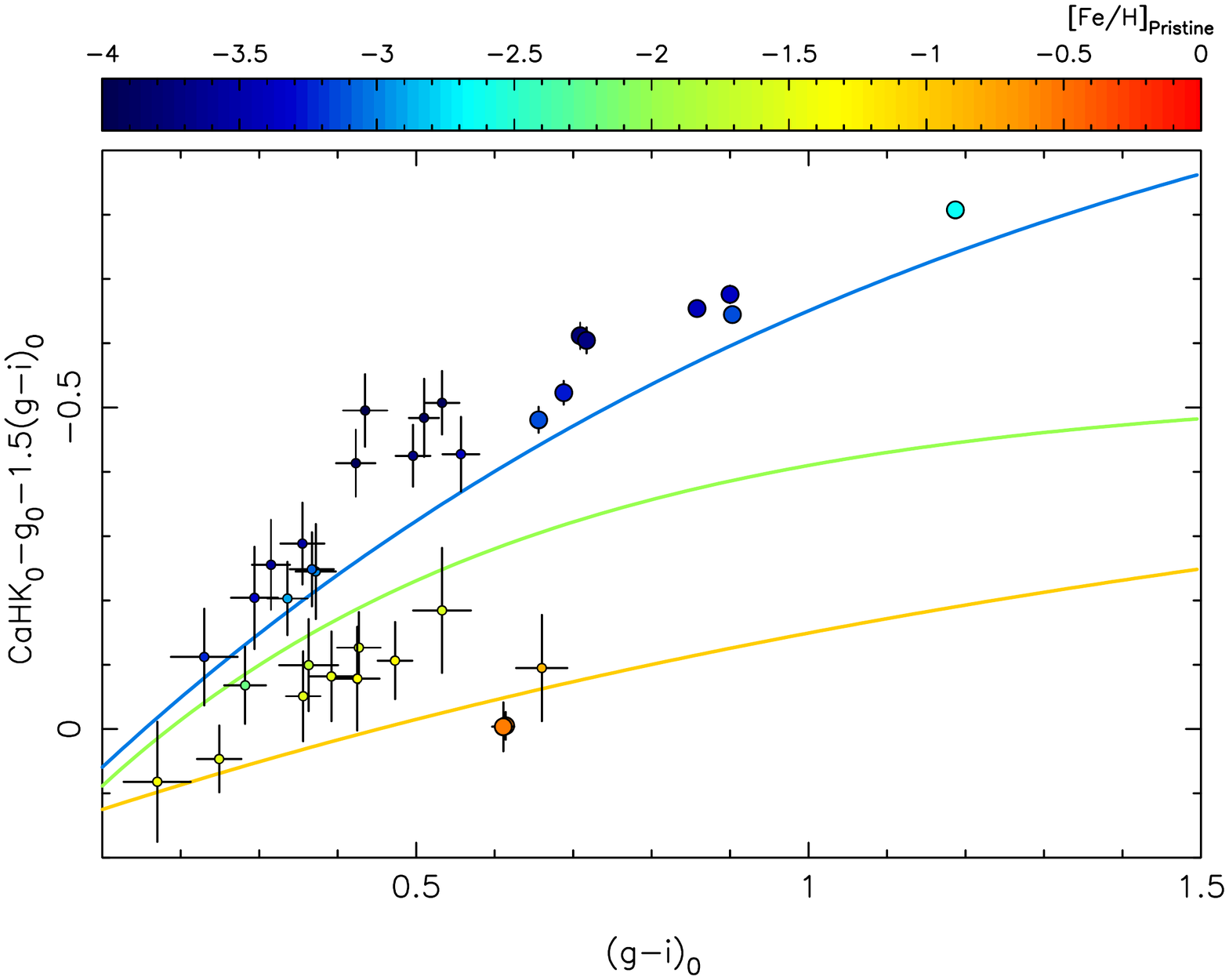}
\caption{\footnotesize Pristine photometric information for all stars of the C-19 stream selected by \texttt{STREAMFINDER} and present in the Pristine survey. Large symbols represent stars with $G_0<19.0$, for which the \texttt{STREAMFINDER} selection is very reliable, and small symbols represent stars fainter than this limit, which are more likely contaminated in the \texttt{STREAMFINDER} catalogue (see also Figure~\ref{CMD}). The lines represent model expectations as determined from the spectral libraries and filter response curves, without assumption on whether the star is a dwarf or a giant\cite{starkenburg17b}. Both model lines and data point are color-coded by their $\FeH_\mathrm{Pristine}$ metallicities. Most C-19 candidate members are located in the region that corresponds to metallicities below $\FeH=-3.0$ (above the blue line). For the large data points, we specifically used a photometric metallicity model tailored to giant stars. Near the tip of the red-giant branch, it deviates significantly from the generic model represented by the colored lines and explains the higher metallicity of the reddest point compared to the models.\label{Pristine_plot}}
\end{center}
\end{figure}

The \emph{Pristine} survey\cite{starkenburg17b} overlaps the region where C-19 stars are found by the \texttt{STREAMFINDER} algorithm, and their combined analysis indicates that C-19 has a very coherent metallicity signal\cite{martin21a} in the extremely low metallicity range ($\FeH_\mathrm{Pristine}<-3.0$), as shown in Extended Data Figure~\ref{Pristine_plot}, although contamination creeps in for faint magnitude ($G_0>19.0$, represented by small symbols in the figure; see also Figure~\ref{CMD}). The mean metallicity of those stars, determined from a sigma-clipping procedure, converges on $\langle[\textrm{Fe}/\textrm{H}]_\mathrm{Pristine}\rangle=-3.43\pm0.07$. From the photometric metallicities measured by \emph{Pristine}, it can already be concluded that C-19 is likely to be the most metal-poor stellar structure ever discovered. \emph{Pristine} metallicities can also be used to flag significantly more metal-rich outliers with $\FeH_\mathrm{Pristine}>-2.5$, as rejected by the sigma-clipping procedure (stars shown in red in Figure~\ref{CMD}).

\section{Mass}
\captionsetup[figure]{labelfont={bf},name={Extended Data Figure}}

We first note that the mass of the C-19 stream that we estimate here from its discovered member stars can only be a lower limit to the true mass of C-19 since it is unlikely that we have discovered the full track of C-19 and its stars may extend much further than detected here. To estimate the mass of C-19, we sum the fluxes of all likely member stars shown in the colour-magnitude diagram of Figure~\ref{CMD}. These correspond to \emph{Pristine}-selected stars with $\FeH_\mathrm{Pristine}<-2.5$, complemented with all stars that share the proper motion of the C-19 stars in this part of the sky and have a Gaia parallax-measurement that is consistent with zero (hollow circles in Figure~\ref{CMD}). All those stars have colour-magnitude properties that are very consistent with the C-19 color-magnitude diagram.

The total observed flux from all these stars with $G_0<18.0$, where we are quite confident the data are complete and suffer from little contamination, is $2.2\times10^3\lsun$. Correcting for fainter stars by using the luminosity function\cite{bressan12} associated to the isochrone shown in Figure~\ref{CMD}, we find that the total luminosity of C-19 is at least $3.9\times10^3\lsun$. Model mass-to-light ratios of very old stellar populations\cite{maraston05} imply $M/L=2$--3 and, consequently, that the mass of C-19 is at least $\sim0.8\times10^4\msun$.

We also use an alternate method to derive the total flux/mass of C-19, relying this time on star counts. We determine the typical luminosity of a system with the age, metallicity, and distance properties of the isochrone displayed in Figure~\ref{CMD} and that hosts the number of stars observed in C-19 above $G_0=18.0$ (18 stars). This analysis yields a slightly smaller average total luminosity ($2.7\times10^3\lsun$) but with a long tail towards higher fluxes that includes the value determined previously.

\section{Heliocentric distance}

For each star that \texttt{STREAMFINDER} highlights as a likely member of a stream, the algorithm determines a most likely heliocentric distance, based on the local distribution of stars on the sky, in proper motion space, and in CMD space. While these distances are not individually precise, they provide, as an ensemble, a first guess for the distance to a stream. In the case of C-19, \texttt{STREAMFINDER} assigns a heliocentric distance between 16 and $22\kpc$ for significant members, with no obvious distance gradient in the region over which C-19 was discovered.

We confirm this rough distance estimate using the blue horizontal branch stars of C-19 that are visible in the CMD shown as triangles in the left-hand panel of Figure~\ref{CMD}. Following equation (7) of Deason et al. (2011)\cite{deason11}, we calculate an average distance of $17.5\kpc$ from these 7 blue horizontal branch (BHB) stars. It should however be noted that the relation between the colour and the absolute magnitude of blue horizontal branch stars has only been calibrated for systems with $\FeH>-2.3$ and that this relation could be biased for C-19 stars that are in the extremely metal-poor regime.

An independent measure of the heliocentric distance is obtained from Gaia parallaxes when those are used to fit for the best C-19 orbit (see below). In this case, the favored orbit has a distance of $20.9\pm0.3\kpc$ ($\pm1\sigma$ Gaussian uncertainties) but relies on parallaxes that are, individually, very small, and could therefore be prone to significant systematic biases, despite taking the average parallax zero-point offset into account\cite{lindegren21}.

As a consequence, we keep the distance constraint loose when estimating the chemical abundances of C-19 stars and check results for distances of 16, 18, and $20\kpc$. A distance of $18\kpc$ is marginally favored to minimize the difference between the FeI and FeII iron abundances. This distance is compatible with the most metal-poor PARSEC isochrone, as shown in Figure~\ref{CMD} and this is the distance we assume for C-19 for the analysis presented in this paper. Future studies and the potential discovery of stars beyond the currently known extent of C-19 may help refine the distance to the stream.

\section{Orbit}
\captionsetup[figure]{labelfont={bf},name={Extended Data Figure}}
\begin{figure}
\begin{center}
\includegraphics[width=1.0\hsize]{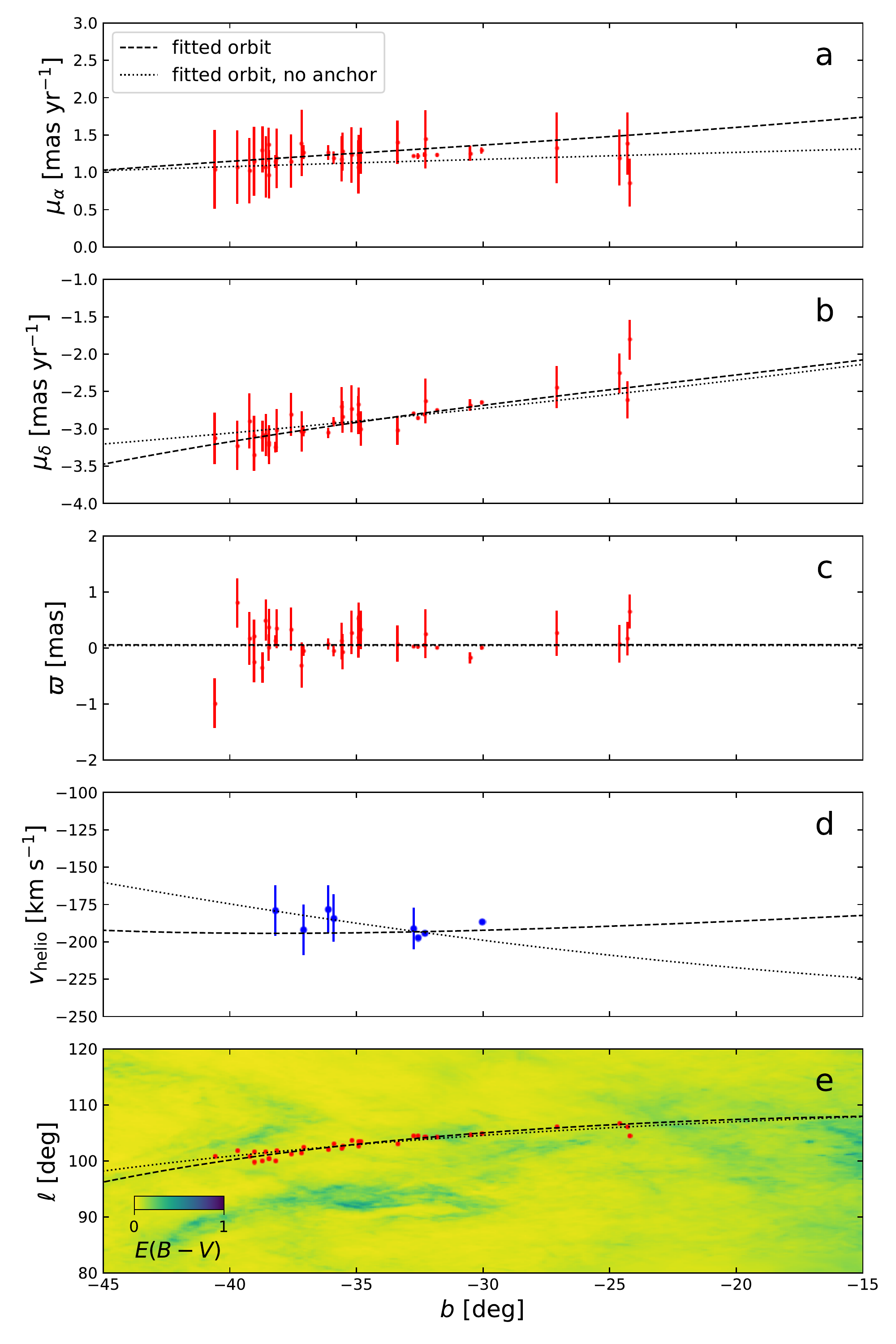}
\includegraphics[width=1.09\hsize]{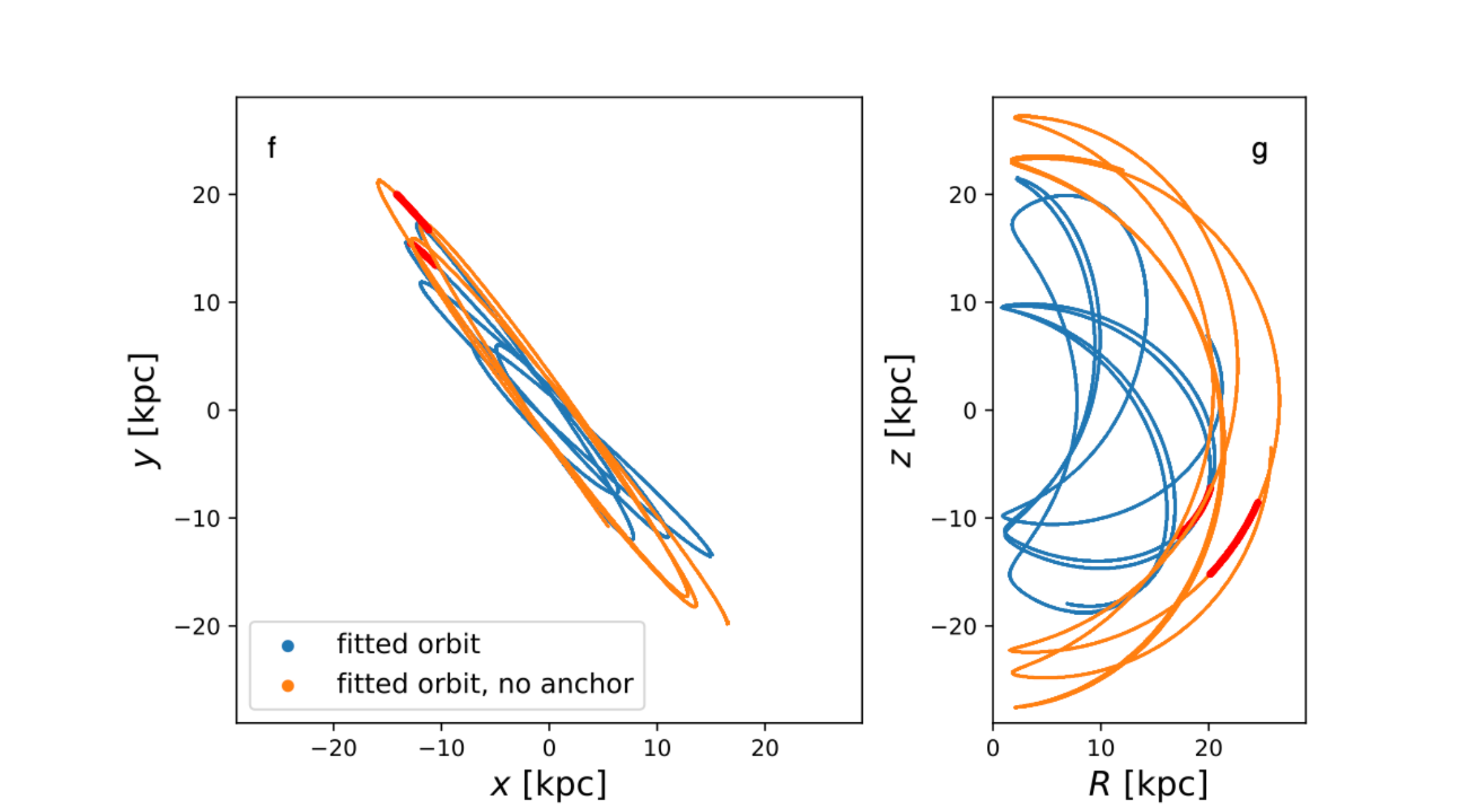}
\caption{\footnotesize Favorite orbital solutions for the C-19 stream. The dashed line shows the orbit of C-19 constrained using the proper motions (red symbols in panels a and b) of C-19 members identified by \texttt{STREAMFINDER}, their Gaia parallax (panel c), their radial velocity when available (panel d) and their location on the sky compared to the distribution of extinction\cite{schlegel98} (panel e). The orbit determined without using the Gaia parallax information but instead anchoring the distance at $18\kpc$ is represented by the dotted line. Panels f and g display the two orbits, intergrated for $\pm1\Gyr$, projected on the Galactic plane, and in the $R$--$z$ plane. The thick red lines correspond to the part of the orbits that overlaps the observed C-19 member stars.\label{orbit}}
\end{center}
\end{figure}

We calculate the orbit of the C-19 stream using the procedure detailed elsewhere\cite{ibata18,ibata21} that has been developed for \texttt{STREAMFINDER} detections. The sample of stars is the same as before: likely members identified by \texttt{STREAMFINDER} with stars that show $\FeH_\mathrm{Pristine}>-2.5$ remove. Extended Data Figure~\ref{orbit} shows the best orbit in two cases: when using the (small) Gaia parallax information of likely stream stars, which results in a distance of $20.9\pm0.3\kpc$ (dashed line in the figure), or when anchoring the distance at $18\kpc$, as favored by the system's BHB stars and the consistency of the FeI vs. FeII abundances (dotted line). Both solutions use the proper motions and spatial locations of all C-19 members identified by \texttt{STREAMFINDER}, along with the radial velocity of the 8 stars for which we obtained spectroscopy (see below). Over the $\sim15\deg$ covered by C-19, these two orbital solutions give very similar results. As shown in panel (d) of Extended Data Figure~\ref{orbit}, determining the velocity gradient of C-19 using more accurate radial velocities will help pin down the distance to the system. 

For both solutions, C-19 is on a remarkably polar orbit that is almost perpendicular to the Galactic plane and remains within the inner confines of the Milky Way with a pericenter of $7.0\kpc$ and an apocenter of $27.4\kpc$ for the distance-anchored orbit ($10.8\kpc$ and $27.4\kpc$, respectively, for the orbital solution constrained by the Gaia parallaxes). The C-19 orbit is slightly prograde and does not share the same plane as any of the growing number of stellar streams on polar orbits (Sagittarius, Cetus, or LMS-1).

\section{Spectroscopy}

Spectra were obtained for eight stars in C-19 and one star that, as expected from the \emph{Pristine} metallicities, turned out to be a field contaminant (see Extended Data Table~\ref{tab:spectraltargets}). This includes high resolution spectra (R$\sim$45,000) for three C-19 stars observed with the Gemini Remote Access to CFHT ESPaDOnS Spectrograph, GRACES\cite{chene14, pazder14}, at the 8.1-m Gemini-North telescope on Maunakea in Hawaii (USA), and medium resolution (R$\sim$2,400) spectra for seven targets (including one observed with GRACES) with the OSIRIS spectrograph mounted on the 10.4-m Gran Telescopio Canarias (GTC) at Roque de los Muchachos Observatory in La Palma (Spain). Examination of the LAMOST database also revealed a good spectrum of the brightest star associated with C-19 (P355.32+27.59), which we also observed with GTC/OSIRIS.

\subsection{GRACES spectra}

The Gemini/GRACES spectra were taken as part of a Large and Long program (GN-2020B-LP-102, PI Venn) between 22 December 2020 and 6 January 2021. The 2-fibre (object+sky) mode was used, and the data initially reduced using the Gemini ``Open-source Pipeline for ESPaDOnS Reduction and Analysis" tool, OPERA\cite{martioli12}. Additional steps were taken to improve the final 1D spectra, including improving the SNR in the overlapping order region by inversely weighing by the noise and improving the continuum normalization using an asymmetric sigma-clipping routine\cite{kielty21}. The full spectral range is $\sim$4,500 \AA\ to $\sim$1 micron; however, light below $\sim$4,800 \AA\ is severely limited by poor transmission through the very long optical fibre. Radial velocities were determined from a cross-correlation with a high SNR Gemini/GRACES spectrum of the metal-poor standard star HD~122563. The target information, exposure times, radial velocities, and the SNR of the final combined spectra are listed in Extended Data Table~\ref{tab:spectraltargets}. 

Effective temperatures were derived from the $G_{BP}-G_{RP}$ colour, corrected for reddening\cite{schlegel98,schlafly11}. We used a colour-temperature calibration based on the IRFM temperatures\cite{gonzalez-hernandez09}, updated for the Gaia photometry EDR3 \cite{mucciarelli21}. Surface gravities were computed from the Stefan-Boltzman equation for an assumed distance of $18 \kpc$ and a stellar mass of $0.8\msun$. Finally, we computed bolometric corrections using a new grid of ATLAS models (Mucciarelli et al., in preparation). Our analysis of 30 similar metal-poor red giants with Gemini/GRACES spectra showed that the typical uncertainties in these stellar parameters are $\pm100$~K in effective temperature and $\pm0.1$ dex in log~gravity\cite{kielty21}. We adopt these uncertainties here as well. Microturbulence was initially determined from the relationship with surface gravity for giants\cite{mashonkina17}; however, raising those initial values slightly ($+0.2 \kms$) helped to flatten the slope observed in the FeI line abundances versus their equivalent width measurements. These stellar parameters are listed in Extended Data Table~\ref{tab:graces}. We note that precision stellar parameters for extremely metal-poor benchmark stars on the red giant branch is currently a significant challenge\cite{karovicova20,giribaldi21,kielty21}. The uncertainties on the stellar parameters and our previous analysis of similar stars and data imply that the quoted $\FeH$ metallicities have systematic uncertainties at the level of $\pm0.2$\,dex.
 
Chemical abundances are determined using the 1D local thermodynamic equilibrium (LTE) ATLAS12 \cite{kurucz05} model atmospheres and MOOG spectrum synthesis code \cite{sneden73,sobeck11}.  The spectral lines examined are listed in Extended Data Table~\ref{tab:glines}, including their equivalent width measurements and atomic data from {\it linemake}\cite{linemake21}.  The abundance uncertainties are estimated from the stellar parameter uncertainties, added in quadrature with an equivalent width measurement uncertainty. The latter are estimated from the standard deviation in the FeI abundances divided by the root of the number of lines of species $X$, i.e., $\sigma$(FeI)/$\sqrt{N_X}$.  

Departures from LTE are known to affect several atomic species (i.e., Na\,I, Ca\,I, Cr\,I, Fe\,I) in metal-poor red giant stars, due to the impact of the stellar radiation field on the statistical populations.  Non-LTE corrections have been determined from the INSPECT\cite{inspect} and MPIA databases\cite{nlte_mpia,lind12,bergemann10,bergemann12,mashonkina07}, and privately by coauthor L. Mashonkina.  These are reported for individual lines in Extended Data Table~\ref{tab:glines}. As all three stars have very similar stellar parameters, the NLTE corrections are similar for all three stars. They are applied in Extended Data Table~\ref{tab:graces}. Only Fe\,II is unaffected by NLTE corrections and we find it is also much less sensitive to uncertainties in the stellar parameters than the other elements (particularly Fe\,I). Calibrations for the extremely metal-poor benchmark stars on the red giant branch \cite{karovicova20} also find Fe\,I is far more sensitive to uncertainties in their stellar parameters. For these reasons, we adopt log(Fe\,II/H) as representative of the metallicity in the three GRACES stars (regardless that it is determined from only 1--2 lines, compared to Fe\,I). All abundances are scaled relative to the solar values\cite{asplund09}, such that [X/Fe] = (log(X/H) $-$ log(Fe/H))$_*$ $-$ (log(X/H) $-$ log(Fe/H))$_\odot$.

\subsection{OSIRIS Spectra}
\captionsetup[figure]{labelfont={bf},name={Extended Data Figure}}
\begin{figure}
\begin{center}
\includegraphics[width=0.55\textwidth,angle=0]{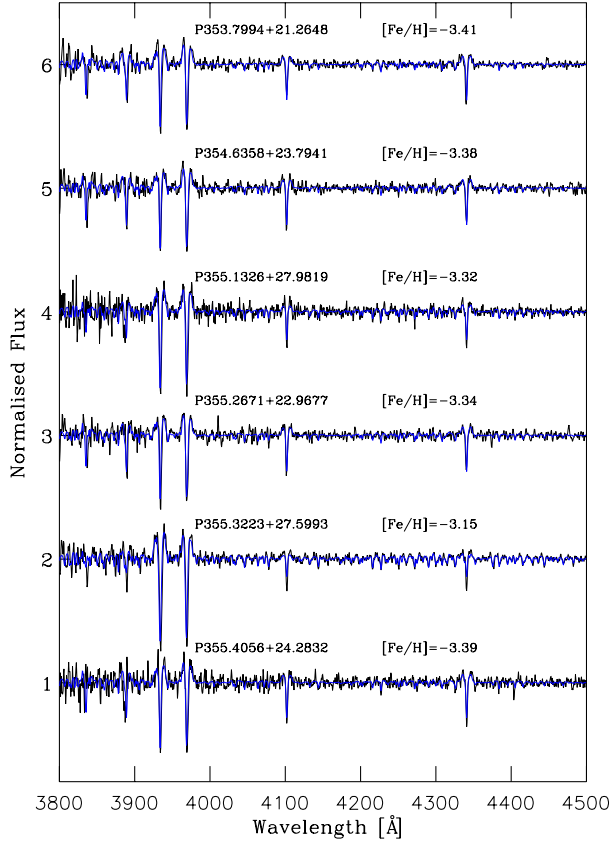}
\caption{\small Spectra of the C-19 member stars observed with OSIRIS, normalised using a running mean filter in the rest frame (black lines), together with the best fit (blue lines) derived by adopting. The metallicity, [Fe/H], computed from [M/H] and [Ca/H] is also indicated for each target (see the text for more details).\label{fig:OSIRIS_forced}}
\end{center}
\end{figure}

Directors Discretionary Time (GTC12-20BDDT) was used to collect spectra for fainter C-19 stars with the OSIRIS spectrograph at the GTC over the period 19--24 January 2021 for a total of $\sim6$\,hours, including overheads. OSIRIS was used in longslit mode, with the 2500U grating to cover a wavelength range 3,440$-$4,610\,\AA, with a $1.0$ arcsec slit and $2\times2$ binning, leading to a resolving power $R\sim2,400$. For the five fainter targets, two observing blocks were used of 1,500\,s each, and only one observing block of 600 and 300\,s were devoted to the two brighter objects (Extended Data Table~\ref{tab:spectraltargets}).  

The data reduction was performed using the {\tt onedspec} package within the IRAF environment\cite{tody93}. The individual spectra were normalized with a third order cubic spline with the {\tt continuum} task in the IRAF environment.  A high-quality spectrum of the extremely metal-poor, bright star G64-12 from previous GTC/OSIRIS observing programs\cite{aguado17,aguado18} was used as a template spectrum to derive the radial velocities of the C-19 candidate stars. We first cross-correlated the brightest target in our sample, Pristine\_355.32+27.59, for which there is also a LAMOST spectrum, with the spectrum of G64-12. This was used to build a cross-correlation function (CCF) with the IRAF task {\tt fxcor} over the spectral range 4,000--4,400 {\AA} to minimise possible distortions of the CCF peak due to interstellar medium (ISM) CaII features. This resulted in a radial velocity of $-191\pm14\kms$ (see Extended Data Table~\ref{tab:spectraltargets}), consistent with the value of $-193.8\kms$ (no uncertainty) from the LAMOST catalogue\cite{li18b}.

Given the similarity in spectral type among the C-19 candidates, the remaining spectra were cross-correlated with that of Pristine\_355.32+27.59. For each star, the weighted mean of the radial velocities from each individual spectrum was calculated along with the corresponding uncertainty. The final radial velocities of all C-19 stars are given in Extended Data Table~\ref{tab:spectraltargets}, with the radial velocity uncertainties computed as the quadratic addition of the radial velocity uncertainty of each star and that of Pristine\_355.32+27.59. We note that the sample also includes an obvious outlier, Pristine\_354.77+32.68, with velocity $-49\pm18\kms$. This star also has a high metallicity measured from the \emph{Pristine} survey ($\FeH_\mathrm{Pristine}\sim-0.7$), confirming that it is not a member of the C-19 structure. The mean radial velocity for the confirmed C-19 members in the OSIRIS sample is $-187.3\kms$ with a standard deviation of $\sim8.0\kms$, consistent with the results for the targets observed with Gemini/GRACES spectra.

Stellar parameters are determined using the same methods as for the Gemini/GRACES sample above; these are shown in Extended Data Table~\ref{table:S1}.  Chemical abundances are determined with the {\tt FORTRAN} \ferre code\cite{allendeprieto14}.  The bluest and noisy parts of the spectra ($\lambda<3,800$\AA) are removed and the spectra are normalized using a running mean filter with a 30 pixel width\cite{aguado17,aguado18}. For the model spectra, a grid of synthetic spectra previously computed\cite{aguado17b} with the {\tt ASS$\epsilon$T} code\cite{koesterke08} is adopted, but enlarged towards lower effective temperatures. Microturbulence ($\xi=2\kms$) and $\alpha$-element abundances ([$\alpha$/Fe]$=+0.4$) are fixed, whereas carbon ranges over a factor of 5 (from [C/M] $= -1.0$ to $+4.0$, where [M/H] represents the global metallicity). It is possible to derive a carbon abundance using the G band near 4300\,\AA\ in these OSIRIS spectra, with [C/Fe] values listed in Extended Data Table~\ref{table:S1}.

The \ferre code is able to look for the best fit by using the Boender-Timmer-Rinnoy Kan global algorithm\cite{boender82} to minimize the solution function. In addition, the overall metallicity, [M/H], derived with a spectral grid that assumes [$\alpha$/M]$=+0.4$, can be refined by deriving a representative $\alpha$ abundance, [Ca/H], based on the two resonance lines Ca H\&K at 3,933 and 3,968\,\AA. We therefore use a \ferre option called FErre Spectral WIndows (FESWI), which applies spectral masks in wavelength ranges with specific chemical information. In this case, we select a spectral window of 50 pixels around the two resonance lines and run \ferre to fit for [Ca/H]. From there, we can derive the iron abundance of each star by simply applying $\rm [Fe/H]=[M/H]-[Ca/M]+0.4$. The uncertainties on the resulting [Fe/H] values are determined from the quadratic sum of the uncertainties on [M/H] and [Ca/H]. Extended Data Figure~\ref{fig:OSIRIS_forced} shows the OSIRIS spectra (black lines) together with the initial fits derived with {\ferre} (blue lines). The mean metallicity of the C-19 members from the OSIRIS sample alone is $\FeH=-3.33$ and the stream has an unresolved metallicity dispersion.

As a sanity check, we also analyze the OSIRIS spectra with FERRE, but in a mode without any assumption on the stellar parameters. While it can be difficult to determine these directly from the spectra of extremely low metallicity stars that, apart from Ca H\&K and Balmer lines, have only relatively weak features, the results yielded by this analysis are on the whole similar to those from the analysis with fixed stellar parameters.

\begin{addendum}
 \item[Acknowledgements] NFM, RI, AA, and ZY gratefully acknowledge support from the French National Research Agency (ANR) funded project ``Pristine'' (ANR-18-CE31-0017) along with funding from CNRS/INSU through the Programme National Galaxies et Cosmologie and through the CNRS grant PICS07708 and from the European Research Council (ERC) under the European Unions Horizon 2020 research and innovation programme (grant agreement No. 834148). KAV is grateful for funding through the National Science and Engineering Research Council Discovery Grants and CREATE programs. ES acknowledges funding through VIDI grant ``Pushing Galactic Archaeology to its limits" (with project number VI.Vidi.193.093) which is funded by the Dutch Research Council (NWO). JIGH acknowledges financial support from the Spanish Ministry of Science and Innovation (MICINN) project AYA2017-86389-P, and also from the Spanish MICINN under 2013 Ram\'on y Cajal program RYC-2013-14875. GT acknowledge support from the Agencia Estatal de Investigaci\'on (AEI) of the Ministerio de Ciencia e Innovaci\'on (MCINN) under grant with reference (FJC2018-037323-I). We gratefully thank the CFHT staff for performing the \emph{Pristine} observations in queue mode, for their reactivity in adapting the schedule, and for answering our questions during the data-reduction process. We are also grateful to the High Performance Computing centre of the Universit\'e de Strasbourg and its staff for a very generous time allocation and for their support over the development of the \texttt{STREAMFINDER} project. 
  
This research has made use of use of the SIMBAD database\cite{wenger00}, operated at CDS, Strasbourg, France (Wenger et al. 2000). This research has made use of the VizieR catalogue access tool\cite{ochsenbein00}, CDS, Strasbourg, France.
 
  Based on observations obtained with MegaPrime/MegaCam, a joint project of CFHT and CEA/DAPNIA, at the Canada-France-Hawaii Telescope (CFHT) which is operated by the National Research Council (NRC) of Canada, the Institut National des Science de l'Univers of the Centre National de la Recherche Scientifique (CNRS) of France, and the University of Hawaii. The observations at the Canada-France-Hawaii Telescope were performed with care and respect from the summit of Maunakea which is a significant cultural and historic site.

This work is based on observations obtained with Gemini Remote Access to CFHT ESPaDOnS Spectrograph (GRACES), as part of the Gemini Large and Long Program, GN-2020B-LP-102.  Gemini Observatory is operated by the Association of Universities for Research in Astronomy, Inc., under a cooperative agreement with the NSF on behalf of the Gemini partnership: the National Science Foundation (United States), the National Research Council (Canada), CONICYT (Chile), Ministerio de Ciencia, Tecnolog\'{i}a e Innovaci\'{o}n Productiva (Argentina), Minist\'{e}rio da Ci\^{e}ncia, Tecnologia e Inova\c{c}\~{a}o (Brazil), and Korea Astronomy and Space Science Institute (Republic of Korea).  CFHT is operated by the National Research Council of Canada, the Institut National des Sciences de l'Univers of the Centre National de la Recherche Scientique of France, and the University of Hawai'i. ESPaDOnS is a collaborative project funded by France (CNRS, MENESR, OMP, LATT), Canada (NSERC), CFHT, and the European Space Agency.  Data was reduced using the CFHT developed OPERA data reduction pipeline.
 
  The authors wish to recognize and acknowledge the very significant cultural role and reverence that the summit of Maunakea has always had within the indigenous Hawaiian community.  We are very fortunate to have had the opportunity to conduct observations from this mountain.

  Based on observations made with the GTC telescope, in the Spanish Observatorio del Roque de los Muchachos of the Instituto de Astrofísica de Canarias, under Director's Discretionary Time. This work is partly based on data obtained with the instrument OSIRIS, built by a Consortium led by the Instituto de Astrof\'isica de Canarias in collaboration with the Instituto de Astronom\'ia of the Universidad Aut\'onoma de M\'exico. OSIRIS was funded by GRANTECAN and the National Plan of Astronomy and Astrophysics of the Spanish Government.

 This work has made use of data from the European Space Agency (ESA) mission {\it Gaia} (\url{https://www.cosmos.esa.int/gaia}), processed by the {\it Gaia} Data Processing and Analysis Consortium (DPAC, \url{https://www.cosmos.esa.int/web/gaia/dpac/consortium}). Funding for the DPAC has been provided by national institutions, in particular the institutions participating in the {\it Gaia} Multilateral Agreement.
  
 Guoshoujing Telescope (the Large Sky Area Multi-Object Fiber Spectroscopic Telescope LAMOST) is a National Major Scientific Project built by the Chinese Academy of Sciences. Funding for the project has been provided by the National Development and Reform Commission. LAMOST is operated and managed by the National Astronomical Observatories, Chinese Academy of Sciences.

\item[Competing Interests] The authors  have no competing financial interests.

\item[Author contributions] NFM is the co-leader of the Pristine survey, led the discovery of the C-19 stream, coordinated the spectroscopic observations and led the writing of the manuscript. KAV led the GRACES spectroscopic follow-up, the analysis of the resulting spectra, and co-led the writing of the manuscript. DSA led the analysis of the OSIRIS spectra and the writing of the corresponding section of the manuscript. ES is the co-leader of the Pristine survey and derived the Pristine photometric metallicities. JIGH coordinated the OSIRIS follow-up, performed the radial velocity analysis of these spectra and was heavily involved in writing this part of the manuscript. RAI led the \texttt{STREAMFINDER} analysis and derived the orbit of the stream. PB, EC, and FS derived stellar and orbital parameters for the stars with spectroscopic follow-up. All other authors helped in the development of the Pristine survey and all authors assisted in the development and writing of the paper. AM derived the relations used to infer the stellar parameters of the spectroscopically observed stars.

\item[Data and code availability] The data used in this paper are listed in Extended Data Tables~\ref{tab:members}, \ref{tab:spectraltargets}, \ref{tab:graces}, \ref{tab:glines} and~\ref{table:S1}. The codes used for the analysis were not designed to be made public but can be requested to the corresponding author.

and code used for the analysis are available upon request to the corresponding author.

\item[Correspondence] Reprints and permissions information is available at www.nature.com/reprints. Correspondence and requests for materials should be addressed to NFM (nicolas.martin@astro.unistra.fr).




\renewcommand{\arraystretch}{0.65}

\begin{table*}
\captionsetup{labelfont={bf},name={Extended Data Table}}
\caption{\small List of potential C-19 member from the \texttt{STREAMFINDER} sample.\label{tab:members}}
\begin{center}
\tiny{
\hspace*{-0cm}\resizebox{1.0\linewidth}{!}{%
\begin{tabular}{lccccccccccccccc}
\hline
Gaia ID             &  RA          &   Dec      & $G_0$    & $\delta_{G}$ & $(BP-RP)_0$ & $\delta_{BP-RP}$ & $\varpi$ & $\delta_\mathrm{\varpi}$ & $\mu_\alpha^*$ & $\delta_{\mu\alpha*}$ & $\mu_\delta$ & $\delta_{\mu\delta}$ & $\FeH_\mathrm{Pristine}^\dagger$ & Note\\
(EDR3) & (ICRS) & (ICRS) &  &  &  &  & ($''$) & ($''$) & ($\textrm{mas\,yr}^{-1}$) & ($\textrm{mas\,yr}^{-1}$) & ($\textrm{mas\,yr}^{-1}$) & ($\textrm{mas\,yr}^{-1}$) &  &\\
\hline
1912706716827846528 &  352.0630548 &   35.6712458 & 19.282 &  0.004 &  0.737 &  0.082 &  0.653 &  0.305 &  0.863 &  0.319 & -1.808 &  0.265 &  --- & \\
1911964791293941760 &  352.7665670 &   34.1962811 & 16.670 &  0.003 &  0.996 &  0.010 &  0.038 &  0.062 &  1.805 &  0.062 & -1.431 &  0.050 & -1.59 & non-member\\
2826426202636892544 &  353.7994502 &   21.2648067 & 17.459 &  0.003 &  0.938 &  0.016 &  0.125 &  0.100 &  1.144 &  0.089 & -3.244 &  0.073 & -3.72 & OSIRIS\\
2826300927030372224 &  353.8760338 &   20.3940474 & 19.067 &  0.004 &  0.792 &  0.044 &  0.210 &  0.297 &  1.163 &  0.329 & -3.347 &  0.218 & -3.90 & \\
2872378543071230464 &  353.8805901 &   32.9557055 & 19.739 &  0.004 &  0.772 &  0.073 & -0.349 &  0.427 &  1.381 &  0.489 & -2.483 &  0.310 & -1.06 & non-member\\
1912600270359141888 &  353.9078460 &   36.1051041 & 19.132 &  0.004 &  0.841 &  0.064 &  0.169 &  0.302 &  1.385 &  0.418 & -2.614 &  0.250 &  ---\\
2826352947674483840 &  353.9983456 &   20.8001111 & 19.223 &  0.004 &  0.756 &  0.056 & -0.352 &  0.272 &  1.305 &  0.310 & -3.100 &  0.208 & -3.91 & \\
2826454824298642560 &  354.1179098 &   21.1058495 & 19.035 &  0.004 &  0.820 &  0.041 &  0.006 &  0.236 &  1.368 &  0.228 & -3.184 &  0.185 & -3.41 & \\
2872357338815387136 &  354.1856575 &   32.7555069 & 19.792 &  0.005 &  0.564 &  0.083 & -0.456 &  0.473 &  1.376 &  0.434 & -2.141 &  0.370 & -1.84 & non-member\\
2826456439206381312 &  354.1924982 &   21.1427394 & 19.388 &  0.004 &  0.623 &  0.056 &  0.369 &  0.333 &  0.974 &  0.326 & -3.211 &  0.260 & -2.75 & \\
2826010071845069056 &  354.3366907 &   19.5521098 & 19.169 &  0.004 &  0.721 &  0.049 & -0.349 &  0.285 &  1.347 &  0.306 & -3.228 &  0.237 & -1.47 & non-member\\
2828149927631546624 &  354.4114566 &   23.8164958 & 19.914 &  0.005 &  0.713 &  0.114 &  1.495 &  0.546 &  1.093 &  0.519 & -2.454 &  0.406 & -1.75 & non-member\\
2827756203686469120 &  354.4165880 &   22.6703814 & 19.405 &  0.004 &  0.658 &  0.076 &  0.271 &  0.395 &  1.275 &  0.392 & -3.372 &  0.276 & -2.36 & non-member\\
2827731670833206272 &  354.4218696 &   22.6125350 & 19.520 &  0.004 &  0.755 &  0.069 & -0.304 &  0.402 &  1.392 &  0.444 & -3.034 &  0.271 & -3.57 & \\
2864770364985943424 &  354.4223451 &   26.6375212 & 19.114 &  0.004 &  0.777 &  0.049 &  0.078 &  0.323 &  1.403 &  0.293 & -3.022 &  0.195 & -3.74 & \\
2871586031407465728 &  354.4571315 &   32.3977802 & 19.668 &  0.004 &  0.844 &  0.062 & -0.106 &  0.367 &  1.245 &  0.387 & -2.793 &  0.302 & -1.13 & non-member\\
2827700678349232128 &  354.5012423 &   22.1880743 & 19.514 &  0.004 &  0.707 &  0.069 &  0.337 &  0.383 &  1.151 &  0.357 & -2.807 &  0.289 & -2.93 & \\
2828240980938817792 &  354.5620499 &   24.3362339 & 19.612 &  0.004 &  0.677 &  0.062 &  0.123 &  0.328 &  1.182 &  0.303 & -2.698 &  0.256 & -3.36 & \\
2871101219796791040 &  354.5752008 &   30.6963267 & 18.497 &  0.003 &  0.878 &  0.028 &  0.315 &  0.186 &  1.222 &  0.184 & -2.542 &  0.137 & -0.69 & non-member\\
2865228517736519168 &  354.5832758 &   27.3150892 & 19.740 &  0.005 &  0.606 &  0.073 & -0.507 &  0.439 &  1.121 &  0.421 & -2.636 &  0.310 & -1.67 & non-member\\
2828159930612041344 &  354.6358014 &   23.7941231 & 17.369 &  0.003 &  0.919 &  0.014 &  0.072 &  0.100 &  1.266 &  0.098 & -3.054 &  0.069 & -3.30 & OSIRIS\\
2879312333975462400 &  354.6573697 &   35.9634777 & 19.391 &  0.004 &  0.853 &  0.061 &  0.073 &  0.338 &  1.196 &  0.375 & -2.258 &  0.265 &  --- & \\
2872775084513320192 &  354.6605429 &   33.3859327 & 19.297 &  0.004 &  0.710 &  0.045 &  0.256 &  0.272 &  1.576 &  0.301 & -2.543 &  0.204 & -1.69 & non-member\\
2864329185945227904 &  354.6617901 &   25.0942850 & 19.311 &  0.004 &  0.738 &  0.046 &  0.530 &  0.284 &  1.233 &  0.268 & -2.668 &  0.220 & -3.04 & \\
2871966806028483584 &  354.7700213 &   32.6894334 & 17.741 &  0.003 &  0.905 &  0.016 &  0.018 &  0.136 &  1.427 &  0.142 & -2.670 &  0.096 & -0.70 & OSIRIS, non-member\\
2868052548930201984 &  354.7701576 &   30.2509844 & 15.489 &  0.003 &  1.081 &  0.006 &  0.010 &  0.041 &  1.294 &  0.043 & -2.648 &  0.029 & -3.38 & GRACES\\
2868021350287794560 &  354.8314453 &   29.7990501 & 17.538 &  0.003 &  0.923 &  0.013 & -0.178 &  0.101 &  1.256 &  0.099 & -2.681 &  0.076 &  --- & \\
2826222621186789376 &  354.8642536 &   20.5574948 & 19.742 &  0.005 &  0.720 &  0.083 &  0.168 &  0.472 &  1.020 &  0.437 & -2.895 &  0.371 &  --- & \\
2852259503209614336 &  354.8710332 &   24.4749727 & 19.141 &  0.003 &  0.805 &  0.041 & -0.065 &  0.313 &  1.276 &  0.254 & -2.840 &  0.212 & -3.90 & \\
2872778829724813824 &  354.8786137 &   33.4186700 & 19.761 &  0.004 &  0.636 &  0.084 &  0.264 &  0.406 &  1.327 &  0.474 & -2.442 &  0.283 & -3.25 & \\
2871914613585750272 &  354.8830318 &   32.5154698 & 19.383 &  0.004 &  0.667 &  0.052 & -0.256 &  0.285 &  1.542 &  0.309 & -2.349 &  0.233 & -2.05 & non-member\\
2823040806334830592 &  354.9002104 &   19.7876803 & 19.406 &  0.004 &  0.747 &  0.066 &  0.097 &  0.360 &  1.245 &  0.343 & -3.273 &  0.267 & -1.40 & non-member\\
2865245186507418496 &  354.9213839 &   27.4455205 & 19.078 &  0.004 &  0.771 &  0.055 &  0.345 &  0.317 &  1.469 &  0.295 & -2.756 &  0.204 & -1.57 & non-member\\
2866151046649496832 &  354.9615142 &   28.4659617 & 13.800 &  0.003 &  1.358 &  0.005 &  0.004 &  0.024 &  1.231 &  0.022 & -2.747 &  0.016 & --- & \\
2864431371806333440 &  355.0588699 &   25.3650219 & 19.346 &  0.004 &  0.727 &  0.054 &  0.417 &  0.347 &  1.496 &  0.290 & -2.759 &  0.225 & -1.68 & non-member\\
2866119435690240256 &  355.0691475 &   28.0094903 & 19.476 &  0.004 &  0.831 &  0.080 &  0.252 &  0.435 &  1.441 &  0.391 & -2.626 &  0.298 &  --- & \\
2865368434887899008 &  355.1326832 &   27.9819597 & 15.052 &  0.003 &  1.095 &  0.005 &  0.055 &  0.032 &  1.236 &  0.029 & -2.804 &  0.022 & -3.14 & GRACE \& OSIRIS\\
2826664453062267776 &  355.1364844 &   21.3505987 & 19.399 &  0.004 &  0.727 &  0.075 &  0.500 &  0.368 &  1.073 &  0.412 & -3.083 &  0.284 & -4.00 & \\
2826881812767837440 &  355.1706311 &   21.8067445 & 19.552 &  0.004 &  0.593 &  0.074 &  0.344 &  0.348 &  1.187 &  0.399 & -3.022 &  0.289 &  -3.13 & \\
2872994922416754944 &  355.2060212 &   34.6340794 & 18.481 &  0.003 &  0.849 &  0.039 &  0.078 &  0.197 &  1.654 &  0.174 & -2.477 &  0.161 & -0.69 & non-member\\
2852422883765640320 &  355.2483495 &   25.3717030 & 19.495 &  0.004 &  0.689 &  0.053 &  0.322 &  0.341 &  1.289 &  0.310 & -2.996 &  0.231 & -3.60 & \\
2822928454287324160 &  355.2549449 &   19.2284211 & 19.599 &  0.005 &  0.855 &  0.082 & -0.986 &  0.446 &  1.037 &  0.527 & -3.128 &  0.342 & --- & \\
2827849249857638656 &  355.2671146 &   22.9677418 & 17.524 &  0.003 &  0.933 &  0.012 & -0.040 &  0.099 &  1.271 &  0.091 & -3.028 &  0.073 & -3.14 & OSIRIS\\
2865256628300500352 &  355.2755507 &   27.7483341 & 15.611 &  0.003 &  1.066 &  0.005 &  0.024 &  0.039 &  1.217 &  0.037 & -2.855 &  0.026 & -3.39 & GRACES\\
2865251577418971392 &  355.3224059 &   27.5993570 & 13.996 &  0.003 &  1.282 &  0.005 &  0.024 &  0.020 &  1.222 &  0.018 & -2.785 &  0.014 & -3.26 & LAMOST \& OSIRIS\\
2852418524375994112 &  355.3553606 &   25.2993957 & 19.615 &  0.004 &  0.823 &  0.061 &  0.201 &  0.377 &  1.073 &  0.357 & -2.814 &  0.254 &  --- & \\
2826217261068446336 &  355.3689432 &   20.9285651 & 19.584 &  0.004 &  0.680 &  0.069 & -0.244 &  0.367 &  1.148 &  0.461 & -3.094 &  0.271 &  --- & \\
2852197625617728384 &  355.4056343 &   24.2832978 & 17.168 &  0.003 &  0.957 &  0.009 & -0.047 &  0.100 &  1.196 &  0.082 & -2.907 &  0.063 & -3.79 & OSIRIS\\
2852399381704345600 &  355.5377628 &   25.0715186 & 19.597 &  0.004 &  0.818 &  0.060 &  0.275 &  0.388 &  1.231 &  0.370 & -2.731 &  0.314 &  --- & \\
2823165364682996992 &  355.6829549 &   20.3234309 & 19.650 &  0.005 &  0.563 &  0.094 &  0.805 &  0.441 &  1.072 &  0.492 & -3.220 &  0.330 &  --- & \\
2827324576652743168 &  354.1805918 &   22.0364990 & 16.945 &  0.003 &  0.060 &  0.009 &  -0.061 &  0.074 &  1.206 &  0.073 & -3.045 &  0.058 &  --- & HB\\
2852215046005045248 &  355.1653168 &   24.4418796 & 16.981 &  0.003 &  0.069 &  0.009 &  0.253 &  0.080 &  3.221 &  1.119 & -3.020 &  0.053 &  --- & HB\\
2851613922380927232 &  359.0995336 &   25.0727130 & 16.706 &  0.003 &  0.094 &  0.009 &  0.216 &  0.079 &  0.216 &  0.079 & -3.168 &  0.051 &  --- & HB\\
2828132236662858240 &  354.5801622 &   23.6484716 & 16.762 &  0.003 &  0.193 &  0.009 &  0.074  &  0.068 &  1.184 &  0.069 & -3.027 &  0.047 &  --- & HB\\
2864750818590938752 &  354.8215599 &   26.4910224 & 16.846 &  0.003 &  0.166 &  0.009 &  0.007 &  0.069 &  1.247 &  0.066 & -2.902 &  0.053 &  --- & HB\\
2867763308651765376 &  355.2655209 &   29.4259271 & 16.949 &  0.003 &  0.184 &  0.009 &  -0.034 &  0.088 &  1.305 &  0.080 & -2.906 &  0.065 &  --- & HB\\
2756806814288972288 &  356.0119714 &    6.4344992 & 16.749 &  0.003 &  0.216 &  0.011 &  -0.056 &  0.075 &  1.308 &  0.080 & -3.170 &  0.061 &  --- & HB\\
\hline
\end{tabular}}}
\end{center}
\footnotesize $^\dagger$ When available the photometric metallicities from the Pristine survey are provided in this column. We note that, by construction\cite{starkenburg17b}, the Pristine photometric metallicities do not go below $\FeH_\mathrm{Pristine}=-4.0$ and that the lessened sensitivity of the Pristine narrow-band in this regime means that, although stars can be flagged quite successfully to have $\FeH< -3.0$, the actual photometric metallicity value is less accurate in this regime than for stars with $\FeH>-3.0$.\\
HB: candidate horizontal branch star\\
non-member: considered a non-member based on the Pristine metallicity ($\FeH_\mathrm{Pristine}>-2.5$) and not used in the analysis.
\end{table*}

\begin{table*}
\captionsetup{labelfont={bf},name={Extended Data Table}}
\caption{\small Summary of observations for the C-19 candidate stars \label{tab:spectraltargets}}
\begin{center}
\hspace*{-0cm}\resizebox{1.0\linewidth}{!}{%
\begin{tabular}{lccccccrrl}
\hline
Name &         RA        &  DEC     & Gaia $G$ & $\FeH_\mathrm{Pristine}$ & $\rm t_{ exp}$ & $\rm S/N$$^{a}$ & $v_{\rm r}$  & Observatory & Comment\\
 &        (deg)        &  (deg)     & (mag)  & &  (s)  &  & $(\kms)$  &  & \\
\hline
Pristine\_354.77+30.25 & 354.7701 & +30.2509 & 15.69 & $-3.38$ & $1\times2,400$ &  50 & $-186.7\pm2.2$ & Gemini/GRACES & \\
Pristine\_355.13+27.98 & 355.1327 & +27.9819 & 15.39 & $-3.12$ & $3\times2,400$ & 100 & $-194.4\pm2.0$ & Gemini/GRACES & \\
                          &                 &                 &      &       & $1\times600$  &  45 & $-198\pm16$ & GTC/OSIRIS & \\
Pristine\_355.27+27.74 & 355.2755 & +27.7483 & 15.82 & $-3.41$ & $2\times2,400$ &  55 & $-197.3\pm2.1$ & Gemini/GRACES & \\
Pristine\_353.79+21.26 & 353.7994 & +21.2648 & 17.63 & $-3.71$ & $2\times1,500$ &  47 & $-179\pm17$ & GTC/OSIRIS & \\
Pristine\_354.63+23.79 & 354.6358 & +23.7941 & 17.49 & $-3.25$ & $2\times1,500$ &  49 & $-178\pm16$ & GTC/OSIRIS & \\
Pristine\_355.26+22.96 & 355.2671 & +22.9677 & 17.60 & $-3.11$ & $2\times1,500$ &  44 & $-192\pm17$ & GTC/OSIRIS & \\
Pristine\_355.32+27.59 & 355.3223 & +27.5993 & 14.19 & $-2.63$ & $1\times300$  &  62 & $-191\pm14$ & GTC/OSIRIS & \\
                       &          &          &       &         & &  40 & $-194$       & LAMOST &Li et al. (2018)  \\
Pristine\_355.40+24.28 & 355.4056 & +24.2832 & 17.28 & $-3.78$ & $2\times1,500$ &  38 & $-184\pm16$ & GTC/OSIRIS & \\
\hline
Pristine\_354.77+32.68 & 354.7700 & +32.6894 & 17.89 & $-0.55$ & $2\times1,500$ &  39 &  $-49\pm18$ & GTC/OSIRIS & non-member \\
\hline
\end{tabular}}
\end{center}
\scriptsize{
$^{a}$ Signal to noise at 600\,nm (Gemini/GRACES) and 420\,nm (GTC/OSIRIS).\\
}
\end{table*}

\begin{table}
\captionsetup{labelfont={bf},name={Extended Data Table}}
\begin{center}
\caption{\small Spectroscopic parameters and 1DLTE chemical abundances for the Gemini/GRACES spectra. The uncertainties of Fe\,I measurements correspond to the line to line scatter in the Fe\,I abundances only, whereas uncertainties on [Fe\,II/H] and [X/Fe] combine the measurement uncertainties (where $\sigma$/${\sqrt N}$) and the uncertainties due to the stellar parameters, added in quadrature.  NLTE corrections are listed in Table~\ref{tab:glines}; the averaged NLTE corrections are applied here, other than Na\,I for which individual line calculations per star were applied.  We adopt the metallicities for these stars from [Fe\,II/H].
 \label{tab:graces}}
\tiny{
\begin{tabular}{lccc}
\hline
Parameter       & Pristine\_354.77+30.25  & Pristine\_355.13+27.98  & Pristine\_355.27+27.74  \\
\hline
 T$_{\rm eff}$ (K)  & $4,928\pm100$    &  $4,881\pm100$    & $4,958\pm100$   \\
 log~g              & $1.83\pm0.10$   &  $1.64\pm0.10$   & $1.89\pm0.10$  \\
 $\xi (\kms)$       & $2.13\pm0.10$   &  $2.20\pm0.10$   & $2.12\pm0.10$  \\
$[$Fe\,I/H$]$           & $-3.21\pm0.17$   & $-3.30\pm0.15$    & $-3.15\pm0.17$ \\
$[$Fe\,I/H$]_\mathrm{NLTE}$    & $-3.03\pm0.17$   & $-3.12\pm0.15$    & $-2.97\pm0.17$  \\
$[$Fe\,II/H$]$          & $-3.42\pm0.12$   & $-3.45\pm0.11$    & $-3.42\pm0.17$  \\
$[$Na\,I/Fe\,II$]$        & $+0.73\pm0.18$   & $+0.77\pm0.17$    & $+0.34\pm0.15$  \\
$[$Na\,I/Fe\,II$]_\mathrm{NLTE}$ & $+0.34\pm0.18$   & $+0.37\pm0.17$    & $+0.05\pm0.15$  \\
$[$Mg\,I/Fe\,II$]$        & $+0.27\pm0.17$   & $+0.33\pm0.14$    & $+0.38\pm0.14$  \\
$[$Mg\,I/Fe\,II$]_\mathrm{NLTE}$ & $+0.33\pm0.17$   & $+0.39\pm0.14$    & $+0.42\pm0.14$  \\
$[$Na\,I/Mg\,I$]$         & $+0.46\pm0.18$   & $+0.44\pm0.16$    & $-0.04\pm0.15$  \\
$[$Na\,I/Mg\,I$]_\mathrm{NLTE}$  & $+0.01\pm0.18$   & $-0.02\pm0.16$    & $-0.39\pm0.15$  \\
$[$Ca\,I/Fe\,II$]$        & $+0.36\pm0.20$   & $+0.36\pm0.08$    & $+0.28\pm0.13$  \\
$[$Ca\,I/Fe\,II$]_\mathrm{NLTE}$ & $+0.56\pm0.20$   & $+0.56\pm0.08$    & $+0.48\pm0.13$  \\
$[$Cr\,I/Fe\,II$]$        & $-0.16\pm0.16$   & $-0.17\pm0.15$    & $-0.05\pm0.15$  \\ 
$[$Ba\,II/Fe\,II$]$       & $<0.52$          & $-0.22\pm0.17$    & $<0.39$         \\
\hline
\end{tabular}}
\end{center}
\end{table}

\renewcommand{\arraystretch}{0.5}

\begin{table*}
\captionsetup{labelfont={bf},name={Extended Data Table}}
\begin{center}
\small{
\caption{\small Spectral lines and atomic data used for the chemical abundances for the Gemini/GRACES spectra. Equivalent widths are in m\AA.  As these three stars have very similar stellar parameters, a single NLTE abundance correction estimate is shown per line from three sources (where NLTE = log(X/H)$_{\rm NLTE} -$ log(X/H)$_{\rm LTE}$).  \label{tab:glines}}
\begin{tabular}{lccrrrrcc}
\hline
Wavel.     & Elem &   $\xi$ & log~gf  & Pristine\_355.1+27.9 & Pristine\_354.7+30.2 & Pristine\_355.2+27.7 & NLTE corr$^{a}$& NLTE corr$^{b}$      \\ 
 (\AA)      &     &   (eV)  &         & (m\AA) & (m\AA) & (m\AA) & (m\AA) & (m\AA) \\ 
\hline
  4,890.755  & FeI &   2.88 & -0.38   & 30    &  w  &   55   & +0.18  & +0.22 \\ 
  4,891.492  & FeI &   2.85 & -0.14   & 40    &  39 &   w    & +0.18  & +0.21 \\ 
  4,918.994  & FeI &   2.86 & -0.37   & n     &  w  &   43   &  --      & +0.22 \\ 
  4,920.502  & FeI &   2.83  &  0.06  & 58    &  60 &   57   & +0.23  & +0.20 \\ 
  5,012.068  & FeI &   0.86 & -2.6    & 50    &  55 &   70   &   --     & +0.18 \\ 
  5,041.072  & FeI &   0.96 & -3.09   & 28    &  30 &   37   &  --      & +0.18 \\ 
  5,041.756  & FeI &   1.48 & -2.20   & 36    &  n  &   45   &   --     & +0.18 \\ 
  5,051.635  & FeI &   0.91 & -2.76   & 55    &  n  &   57   &   --     & +0.18 \\ 
  5,083.339  & FeI &   0.96 & -2.84   & 40    &  w  &   33   &  --      & +0.18 \\ 
  5,123.700  & FeI &   1.01 & -3.06   & 14    &  w  &   23   &   --     & +0.18 \\ 
  5,171.596  & FeI &   1.48 & -1.72   & 54    &  58 &   70   &   --     & +0.18 \\ 
  5,194.942  & FeI &   1.56 & -2.02   & 44    &  n  &   n    & +0.20  & +0.18 \\ 
  5,202.336  & FeI &   2.17 & -1.87   & 16    &  n  &   w    & +0.20  & +0.19 \\ 
  5,216.274  & FeI &   1.61 & -2.08   & 27    &  n  &   n    & +0.20  & +0.18 \\ 
  5,227.190  & FeI &   1.56 & -1.23   & 94    &  90 &   80   &  --      & +0.14 \\ 
  5,232.940  & FeI &   2.94 & -0.19   & 51    &  48 &   58   & +0.20  & +0.22 \\ 
  5,266.555  & FeI &   2.99 & -0.49   & 42    &  w  &   w    &   --     & +0.22 \\ 
  5,269.537  & FeI &   0.86 & -1.33   & 135   & 120 &  129   & +0.10  & +0.08 \\ 
  5,281.700  & FeI &   3.04 & -1.02   & 10    &  w  &   w    & +0.19  & +0.21 \\ 
  5,324.179  & FeI &   3.21 & -0.11   & 32    &  b  &   n    & +0.21  & +0.25 \\ 
  5,328.039  & FeI &   0.91 & -1.47   & 112   & 109 &  113   &  --      & +0.10 \\ 
  5,328.532  & FeI &   1.56 & -1.85   & 51    &  37 &   38   &  --      & +0.18 \\ 
  5,371.489  & FeI &   0.96 & -1.64   & 102   & 112 &  112   &   --     & +0.12 \\ 
  5,397.128  & FeI &   0.91 & -1.98   & 91    &  77 &   93   & +0.17  & +0.15 \\ 
  5,405.775  & FeI &   0.99 & -1.85   & 99    & 105 &   96   & +0.16  & +0.14 \\ 
  5,429.696  & FeI &   0.96 & -1.88   & 100   &  93 &  101   &   --     & +0.14 \\ 
  5,434.524  & FeI &   1.01 & -2.13   & 70    &  71 &   60   & +0.20  & +0.16 \\ 
  5,446.917  & FeI &   0.99 & -1.91   & 96    & 105 &   91   & +0.15  & +0.15 \\ 
  5,455.609  & FeI &   1.01 & -2.09   & 77    &  81 &   74   &   --     & +0.16 \\ 
  5,497.516  & FeI &   1.01 & -2.83   & 34    &  25 &   22   &  --      & +0.18 \\ 
  5,501.465  & FeI &   0.96 & -3.05   & 30    &  w  &   w    &   --     & +0.18 \\ 
  5,506.779  & FeI &   0.99 & -2.79   & 33    &  36 &   34   &    --    & +0.18 \\ 
  5,572.842  & FeI &   3.39 & -0.28   & 19    &  n  &   n    & +0.20  & +0.24 \\ 
  5,586.756  & FeI &   3.37 & -0.11   & 23    &  34 &   27   & +0.20  & +0.25 \\ 
  6,136.615  & FeI &   2.45 & -1.40   & 16    &  w  &   n    & +0.18  & +0.21 \\ 
  6,137.691  & FeI &   2.59 & -1.40   & 11    &  w  &   n    & +0.17  & +0.20 \\ 
  6,191.558  & FeI &   2.43 & -1.60   & 18    &  n  &   n    & +0.20  & +0.21 \\ 
  6,230.722  & FeI &   2.56 & -1.28   & 22    &  27 &   n    & +0.20  & +0.20 \\ 
  6,252.555  & FeI &   2.40 & -1.69   & 11    &  w  &   n    & +0.20  & +0.20 \\ 
  4,923.922  & FeII &  2.89 & -1.21   & 73    &  70 &   n    & -0.01  & -0.01 \\ 
  5,018.435  & FeII &  2.89 & -1.35   & 59    &  56 &   58   & -0.00  & -0.01 \\ 
  5,889.951  & NaI &   0.00 &  0.12   & 152   & 135 &  126   & -0.33  & -0.40 \\ 
  5,895.924  & NaI &   0.00 & -0.18   & 137   & 145 &  106   & -0.33  & -0.33 \\ 
  5,172.684  & MgI &   2.71 & -0.40   & 133   & 133 &  145   & +0.08  & +0.02 \\ 
  5,183.604  & MgI &   2.72 & -0.18   & 157   & 143 &  150   & +0.06  & -0.04 \\ 
  5,528.405  & MgI &   4.34 & -0.62   & 20    &  21 &   20   & +0.11  & +0.15 \\ 
  6,102.723  & CaI &   1.88 & -0.89   & 11    &   n &   n    & +0.27  & +0.14 \\ 
  6,122.217  & CaI &   1.88 & -0.41   & 20    &  25 &   b    & +0.25  & +0.14 \\ 
  6,162.173  & CaI &   1.90 &  0.10   & 32    &  32 &   32   & +0.24  & +0.14 \\ 
  6,439.075  & CaI &   2.52 &  0.47   & 20    &  24 &   23   & +0.21  & +0.20 \\ 
  5,204.498  & CrI &   0.94 & -0.19   & 53    &  27 &   50   & +0.56  &  --     \\ 
  5,206.023  & CrI &   0.94 &  0.02   & 39    &  48 &   45   & +0.56  &   --    \\ 
  5,208.409  & CrI &   0.94 &  0.17   & 33    &  48 &   41   & +0.55  &  --     \\ 
  4,934.100  & BaII &  0.00  & -1.16  & 26  & $<$40 & $<$35  & --       & +0.20 \\ 
\hline
\end{tabular}}
\end{center}
\scriptsize{
Note that when lines could not be measured in all three stars, they are noted as
(n, w, b), which refers to (noisy, weak, blended).   \\ 
$^{a}$ NLTE correction estimates from the INSPECT (http://www.inspect-stars.com) and/or 
MPIA database (http://nlte.mpia.de). \\
$^{b}$ NLTE corrections from the private calculations of coauthor L. Mashonkina\cite{mashonkina_NLTE}. Only the Na\,I NLTE corrections show some star to star variations due to the differences in the line strengths (the average from the three stars is reported here).}
\end{table*}

\renewcommand{\arraystretch}{0.65}

\begin{table*}
\captionsetup{labelfont={bf},name={Extended Data Table}}
\caption{\footnotesize Stellar parameters and abundances of C-19 stars observed with GTC/OSIRIS.\label{table:S1}}
\begin{center}
\label{S1}
\hspace*{-0cm}\resizebox{1.0\linewidth}{!}{%
\begin{tabular}{lcccccc}
\hline
Name &  $T_{\rm eff}$ & $\log g$ & $\left[{\rm M/H}\right]$ & $\left[{\rm Ca/H}\right]$ & $\left[{\rm Fe/H}\right]$ & $\left[{\rm C/Fe}\right]$ \\
     &  (K)  &$\rm (cm\,s^{-2})$&&&&               \\
\hline
With stellar parameters based on Gaia photometry (used in the paper):\\
\hline
Pristine\_353.79+21.26 &  $5,230\pm100$ & $2.73\pm0.10$ & $-3.51\pm0.15$ & $-3.21\pm0.11$ & $-3.41\pm0.19$ & $+0.48\pm0.32$   \\         
Pristine\_354.63+23.79 &  $5,280\pm100$ & $2.71\pm0.10$ & $-3.49\pm0.17$ & $-3.20\pm0.10$ & $-3.38\pm0.20$ & $+0.61\pm0.63$     \\     
Pristine\_355.13+27.98 &  $4,881\pm100$ & $1.64\pm0.10$ & $-3.37\pm0.14$ & $-3.02\pm0.12$ & $-3.32\pm0.18$ & $-0.11\pm0.82$     \\      
Pristine\_355.26+22.96 &  $5,248\pm100$ & $2.76\pm0.10$ & $-3.45\pm0.16$ & $-3.17\pm0.11$ & $-3.34\pm0.19$ & $+0.44\pm0.36$     \\      
Pristine\_355.32+27.59 &  $4,570\pm100$ & $1.07\pm0.10$ & $-3.23\pm0.19$ & $-2.92\pm0.10$ & $-3.15\pm0.18$ & $-0.60\pm0.46$    \\      
Pristine\_355.40+24.28 &  $5,193\pm100$ & $2.60\pm0.10$ & $-3.45\pm0.15$ & $-3.11\pm0.11$ & $-3.39\pm0.18$ & $+0.16\pm0.23$   \\      
\hline
Pure spectroscopic stellar parameters:\\
\hline
Pristine\_353.79+21.26 &  $5,319\pm180$ & 1$\downarrow$ & $-3.50\pm0.12$ & -- & -- & --   \\         
Pristine\_354.63+23.79 &  $5,253\pm340$ & 1$\downarrow$ & $-3.56\pm0.15$ & -- & -- & --     \\     
Pristine\_355.13+27.98 &  $4,874\pm105$ & 1$\downarrow$ & $-3.37\pm0.12$ & -- & -- & --     \\      
Pristine\_355.26+22.96 &  $5,290\pm186$ & 1$\downarrow$ & $-3.48\pm0.14$ & -- & -- & --     \\      
Pristine\_355.32+27.59 &  $5,089\pm106$ & 1$\downarrow$ & $-3.02\pm0.18$ & -- & -- & --    \\      
Pristine\_355.40+24.28 &  $5,344\pm142$ & 5$\uparrow$ & $-3.61\pm0.11$ & -- & -- & --   \\      
\hline
Up and down arrows for $\log g$ values represent cases where FERRE reached the limit of the spectral model grid.

\end{tabular}}
\end{center}
\end{table*}

\end{addendum}


\end{document}